\documentstyle[aps,amssymb,psfig,twocolumn]{revtex}
\draft
\def\diff{{\rm d}\,}
\def\exp{{\rm e}}

\begin{document}
\wideabs
{
\title{\LARGE 
{\sf  Dynamics of a non homogeneously coarse grained system}}
\author{
Stefano Curtarolo$^{1}$, and Gerbrand Ceder$^{1,2}$} 
\address{
$^1$Department of Materials Science and Engineering, MIT, Cambridge, MA 02139 \\
$^2$Center for Materials Science and Engineering, MIT, Cambridge, MA 02139 \\
}
\date{\today}
\maketitle
\begin{abstract}
To study materials phenomena simultaneously at various length scales, descriptions
in which matter can be coarse grained to arbitrary levels, are necessary.
Attempts to do this in the static regime (i.e. zero temperature) have already
been developed. In this letter, we present an approach that leads to a dynamics 
for such coarse-grained models.  This allows us to obtain temperature-dependent and transport
properties.  Renormalization group theory is used to create new local
potentials model between nodes, within the approximation of local
thermodynamical equilibrium.  Assuming that these potentials give an
averaged description of node dynamics, we calculate thermal and mechanical
properties. If this method can be sufficiently generalized 
it may form the basis of a Molecular Dynamics method with time and spatial 
coarse-graining.
\end{abstract}
\pacs{PACS numbers: 83.10.-y, 62.20.-x, 81.40.Lm, 82.20.Wt, 63.10.+a}
}
Predicting macroscopic properties of materials starting from an atomistic 
or electronic level description can be a formidable task due to the many orders of
magnitude in length and time scale that need to be spanned. A characteristic of successful
approaches to this problem is the systematic coarse-graining of less relevant degrees
of freedom in order to obtain Hamiltonians that span larger length and time scales. 
For example, in first - principles thermodynamics of crystalline solids, which is one of the best
developed examples of micro to macro bridging, electronic and vibrational excitations
are integrated out in order to obtain lattice - model Hamiltonians that describe the substitutional
degree of freedom \cite{ceder.ref.12}. 
Monte Carlo simulations can then be used to simulate the kinetic evolution of the 
system or to obtain its thermodynamics state function. 
The reason first - principles thermodynamics has been so well developed is that 
it deals with extensive (averaged) quantities of homogeneous materials and minor
inhomogeneities in real materials, such as interfaces or dislocations, have a minor 
effect on the thermodynamic functions.
Another extreme is the study of mechanical properties, such as plasticity, where the property
of interest (i.e. plastic yield) is determined by discrete events (slip of dislocations), but
over a very large scale. 
This type of problems requires the use of inhomogeneous coarse - graining methods:
atomistic - level resolution may be required near the key features in the material 
(e. g. dislocations or grain boundary) and lower resolution is needed in between, in order
to make the problem computationally tractable. An important step towards such a 
coupled multi - length scale description was taken with the development of the
quasicontinuum method (QCM) in which the behaviour of groups of atoms (nodes) are
treated with a finite element scheme \cite{ref.quasi.1}.
The interaction between atoms is typically calculated with empirical potentials.
In its original form QCM is essentially a method that improves the boundary conditions on 
atomistic regions and allows the boundary conditions of different atomistic regions to interact.
Since QCM consists of an optimization of the energy, no time or temperature phenomena are present.
Introducing temperature into QCM can be done by using potentials that incorporate the entropy 
due to lattice vibrations, for example in a local Einstein description 
\cite{ref.quasi.6,ref.einstein.vibration.1}.
This is conceptually similar to the coarse - graining of vibrations in first - principles 
thermodynamics \cite{ceder.ref.3} and leads to structures that are free - energy minimized.
The study of time/temperature dependent phenomena is more difficult and requires the development
of a dynamics for a system with an inhomogeneous level of coarse graining.
To our knowledge, no formal development of this problem exists 
for a more general class of interactions, though methods that couple Molecular dynamics to continuum description 
have been studied  \cite{continuos.coupling.2} and Finite Elements dynamics 
has been developed in the harmonic approximation without time rescaling \cite{ref.rudd}.
Other coarse-graining techniques have been developed: 
effective Langevin dynamics procedures have been proposed to describe relaxation of 
macroscopic degrees of freedom \cite{mori.zwanzig}, and 
transition state theories have been generalized to boost time evolution, 
by modifying the shape of the surface of contant energy 
(hyperdynamics) \cite{voter} or using the Onsager-Machlup action \cite{elber}.
None of these methods has employed simultaneous space/time coarse-graining.

In this letter we present a suggestion for the dynamical modeling of 2D systems that exist 
simultaneously at different level of coarse graining in time and space domains.
Coarse graining requires both a scheme to remove atoms, and a prescription
to define potentials between the remaining atoms. For the approach advocated in this 
letter atoms are integrated out through bond moving, similar to the 
Migdal-Kadanoff approach in renormalization group \cite{ref.mk}.
New potentials can be defined in various ways, but an important aspect is
that the coarse-grained system ultimately evolves to the same equilibrium state of the 
fully atomistic one. Hence, our criterium for defining new potentials when removing atoms
is that the partial partition function of the system remains unchanged.
We assess the validity
of the model, by comparing the elastic, thermodynamic, and transport properties of a
non-homogeneously coarse - grained and fully atomistic model. 

As a matter of introduction, the coarse-graining is first considered for a one-dimensional system.
Let us consider a finite one - dimensional chain of atoms with mass $m$ that interact through
nearest neighbor pair potentials. 
Each particle has kinetic and potential energy and the total Hamiltonian is:
$H[\{ q_i, p_i\}]$, where $q_i$ 
is the position coordinate and $p_i$ is the momentum.
It will be assumed that the gradient of temperature along the chain is small enough for
local thermodynamic equilibrium to exist. 
One coarse graining step is defined as removing every second atom. 
The energy $H_{(1)}(q_i,q_{i+2})$ between nearest neighbors of the coarse system (2nd neighbors 
in the full atomic system) is defined so as to conserve the partition function
(the subscripts $_{(1)}$ indicate the 1st step of renormalization):

{{$$
\!\!\!\!\exp^{-\beta H_{(1)}(q_i,q_{i+2})}\!=\!\frac{1}{h} \!
\int \!\!\!\!\!\int \!\!\diff q_{i+1} \diff p_{i+1}\,
\exp^{-\beta \left[H(q_i,q_{i+1})+H(q_{i+1},q_{i+2})\right]}.
\nonumber
$$}}
Integrating the momenta leads to the the reduced formula:
{{
$$
\label{equation.rg.2}
\!\!\!\!\!\!\exp^{-\beta \left[V_{(1)}(q_i,q_{i+2},T)+\tilde{F}_{(1)}(T,i+1)\right]}\!=\!\!\!\int
\!\!\diff q_{i+1} \, \exp^{-\beta \left[V(q_i,q_{i+1})+V(q_{i+1},q_{i+2})\right]},
$$
}}
where $\tilde{F}_{(1)}(T,i+1)$ is an excess free energy that does not depend 
on the positions $(q_i,q_{i+2})$.
It contains the entropy of atom $i+1$ lost in the renormalization step.
It is necessary to keep track of this quantity to calculate properly the extensive thermodynamic 
quantities of the system. 
Finally, from the last equation, it is possible to extract an effective 
potential $V_{(1)}(q_i,q_{i+2},T)$ (with harmonic potentials the linear chain model can be solved analytically). 
This potential is now temperature dependent \cite{ref.pot.1}.

It is not obvious that this choice for the coarse-graining algorithm leads to 
the correct dynamics. However is is well know that the interaction so defined between
particles $i$ and $i+2$ is correct in the long time limit, i.e. when the motion of
$i+1$ is much faster that $i$ and $i+2$. Hence low frequency dynamics in the coarse system
will likely be better represented than high frequencies. 
However, we feel that the ultimate justification 
for this approach should be evaluated on the basis of a comparison of properties of the coarse and fully
atomistic system. This is investigated in this letter. Hence this potential generates some 
dynamics of particles $i$ and ${i+2}$ by the averaged interaction 
of particle ${i+1}$.
In a first approximation, we consider $\tilde{F}_{(1)}(T,i+1)$ to be independent 
from $(q_i,q_{i+2})$, and the potential $V_{(1)}$ to contain all the possible spatial dependencies 
of the remaining coordinates.

\begin{figure}
\psfig{file=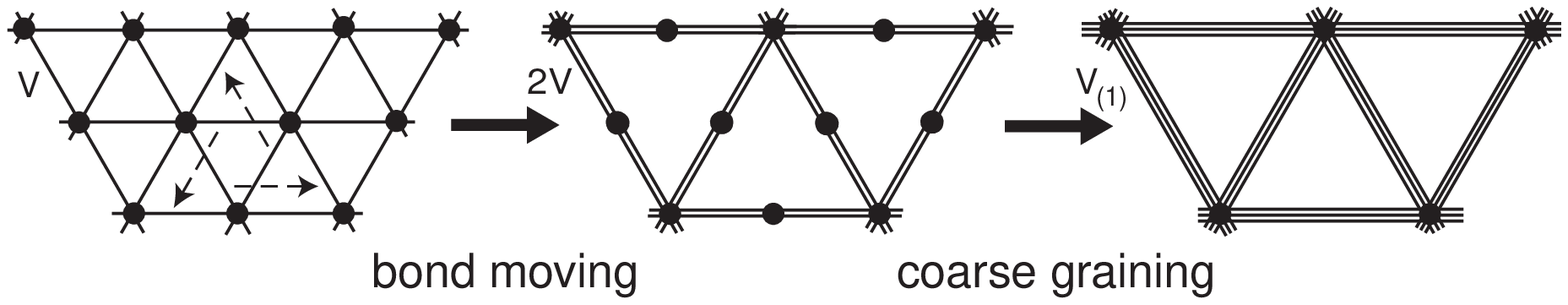,width=8.5cm,clip=}
\vspace{3mm}
\psfig{file=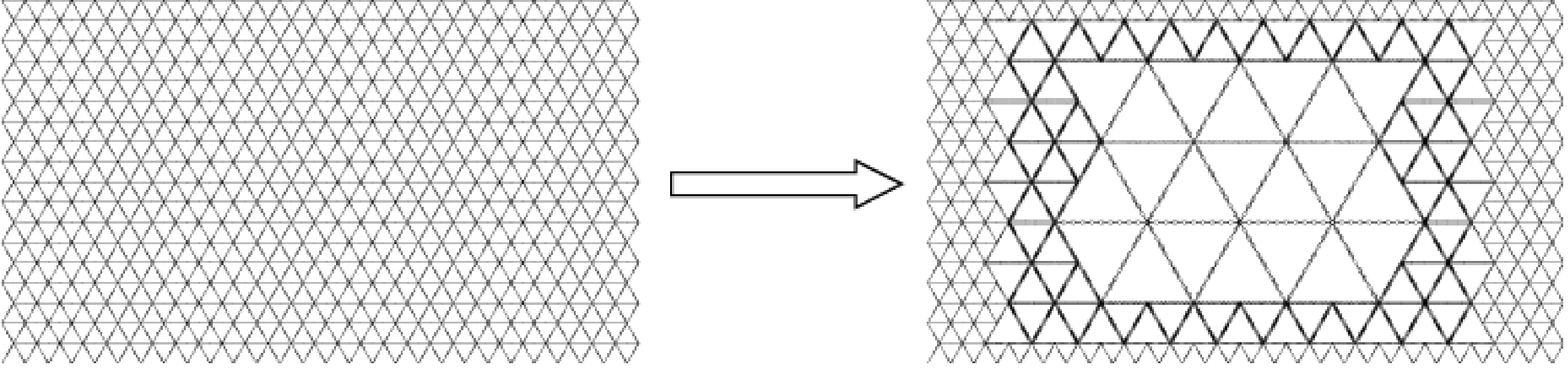,width=8.5cm,angle=-90}
\vspace{1mm}
\caption{Triangular lattice with bond moving approximation, and example of non-homogeneous coarse graining}
\label{label.fig.1}
\end{figure}

\begin{figure}
\psfig{file=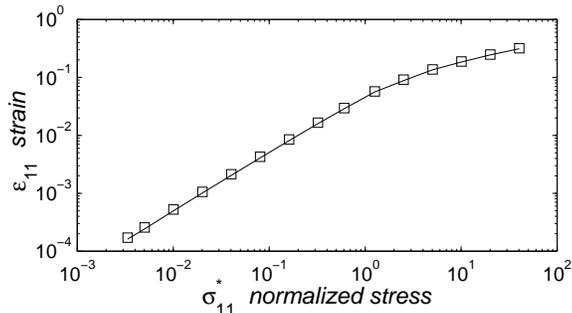,width=77mm,angle=90}
\caption{Strain versus normalized stress $\sigma^\star_{11}$.
        Continuous line represents the strain ${\epsilon_{h}}_{11}$ for the original system,
 	while squares $\Box$ represent the strain ${\epsilon_{nh}}_{11}$ for 
        the non-homogeneously coarse grained one ($T^\star=0.32$)}
       \label{label.fig.2}
\end{figure}

As is typical in dynamic renormalization group theories, integrating out degrees
of freedom, leads to a time rescaling as $t_{(1)}=b^z t$ where $b=2$ is our 
scaling factor and $z$ is the dynamical exponent \cite{ref.drg}.
This exponent does not affect equilibrium properties and will determined later.
Coarse graining has to conserve the total mass, so, we take, 
$m_{i}^{(1)}=m_{i} + m_{i+1}/2$ and 
$m_{i+2}^{(1)}=m_{i+2} + m_{i+1}/2$.
To describe the potential we take a simple expansion around the minimum 
$a$ ($a=3$\AA\ atomistic lattice spacing and $m_i=30\times$neutron mass), 
$V(q_i,q_{i+1})=\sum_n k_n (q_i-q_{i+1}-\alpha_n a)^n$. 
A symmetric potential has $\alpha_n=1,\, \forall n$. 
If we assume that the potential $V_{(1)}$  has the same functional shape of 
$V$, we get recursive relations 
$V_{(j-1)} \stackrel{RG}{\Rightarrow} V_{(j)}$ (with $V_{(0)}\equiv V$) which 
creates higher order coarse-grained potentials: 
{{$$ V_{(j)}(q_i,q_{i+b^j},T)=\sum_n {k_{(j)}}_n (T)(q_i-q_{i+b^j}-b^j {\alpha_{(j)}}_{n} a)^n .  \nonumber
\label{equation.rg.4} $$}}
Our model system interacts through a potential with second and fourth powers 
$k_2= 8.8\cdot10^3$ [K/\AA$^2$], $k_4= 1.8\cdot10^5$ [K/\AA$^4$], which is either
symmetric ($\alpha_2=\alpha_4=1$), or non-symmetric ($\alpha_2=0.999, \alpha_4=1.0315$), 
and we keep only the second and fourth power coefficients 
${k_{(j)}}_2,{k_{(j)}}_4,{\alpha_{(j)}}_{2},{\alpha_{(j)}}_{4}$ 
for every renormalized potential. \\
To extend the method to a 2D lattice, we use the Migdal-Kadanoff moving bonds (fig.\ref{label.fig.1}) 
approximation \cite{ref.mk}, to remove atoms.
The model and assumptions are tested by comparing the results of molecular dynamics simulations
on the original fully atomic system and the non-homogeneously coarse grained one. 
Regions with different coarse graining are considered with different time-evolution, 
because of the dynamical exponent $z$ in the time scaling.
The method is tested on {\it elastic properties}, {\it thermodynamic quantities}, {\it thermal expansion}, 
and finally on its {\it heat transport}.
We see these as successfully more stringent tests to pass.
A 2D triangular lattice with $225 \times 31$ atoms (6975 atoms) and symmetric potential, is 
studied in response to static tensile/shear stress, and under isotropic pressure deformation.
The non-homogeneously coarse grained system is represented by 1510
nodes in an arrangement similar to that in figure \ref{label.fig.1}.
We calculate the strain response to a normalized tensile stress 
$(\sigma^\star_{11}=\sigma_{11}/\sigma_0)$ along the longest direction ($\sigma_0$ 
is the tensile stress which gives $5\%$ of lattice distortion).
Various simulations are done at different normalized temperatures $(T^\star=T/T_0 = 0.04 \cdots 0.64$, where
$T_0={\hbar \omega_0}/{k_B}$, $\omega_0=\sqrt{{k_2}/{m}}$).
Figure \ref{label.fig.2} shows the strain response for $T^\star=0.32$.
The linear / non-linear regions of the strains, 
are clearly conserved by the non-homogeneous coarse graining. 
The elastic modulus of the two systems are equal to within $4\%$.
Similar results have been obtained for shear strain ($5\%$). 
For the bulk modulus calculations we use the non-symmetric potential, as described in the previous section.
In the range of temperatures considered the bulk modulus $B_{h}$ for the original system (subscript $_{h}$) 
and $B_{nh}$ for the non-homogeneous coarse grained one (subscript $_{nh}$) are equal to within $\pm 1\%$
The good agreement for the elastic properties may not be surprising but indicates 
that the bond-folding does not modify the macroscopic energies. 
As the elastic properties are largely a reflection of the 
direct interaction and are not much influenced by temperatures or atomic motion, they do not really test 
the assumption made on the coarse-grained dynamics.
We also evaluate the heat capacity for the 2D system with non-symmetric potential.
The heat capacity from the coarse system can not be directly 
compared to the complete system, but needs to be augmented with
the contribution from the entropy that is lost by removing degrees
of freedom.
This entropy can be calculated by keeping track of all the free energies 
$\tilde{F}_{(j)}(T,i)$ produced by the renormalization
integration and taking temperature derivatives to get entropies \cite{ref.yeomans}.
We find that within the range of simulations 
$T^\star=0.1 \cdots 1.6$, ${C_V}_{nh}$ and ${C_V}_{h}$ are equal to within the numerical noise of $\pm 0.2\%$. 
Since the calculation is classical, the specific heats do not go to zero for 
$T^\star \rightarrow 0$.

The calculation of properties at constant pressure, such as thermal expansion or $C_{p}$, requires
particular effort to reproduce the correct dependence of free energy on temperature and volume.
This dependence is hidden in either the nonlinear terms of the renormalized potentials 
or the volume dependence of the integrated free energy $\tilde{F}_{(j)}(T,i)$.
An approximate representation of the thermal expansion can be achieved by interpolation of 
${\alpha_{(j)}}_{2},{\alpha_{(j)}}_{4}$ to conserve the thermalized bond length 
$
  <q_{i+b^j}-q_i>_{(j)} = b <q_{i+{(b-1)}^j}-q_i>_{(j-1)},
$
(the subscripts on the averages indicate the potential used for the calculation).
This constrained can be satisfied with a very small shift ($\approx 10^{-5} \sim 10^{-4}$)
of the parameters ${\alpha_{(j)}}_{2},{\alpha_{(j)}}_{4}$.
The overall effect is simply an effective bias on the potential 
to match the correct thermal expansion.
Simulations with this method for the thermal expansion coincide for the homogeneous and 
the non homogeneous coarse grained systems, within the numerical noise ($\pm 5\%$).
It is important to understand that without this correction, coarse-grained systems
have incorrect thermal expansion (due to the removal of volume-dependent entropy).
Hence simulations with {\it inhomogeneously} coarse grained regions would build up large internal
strains upon changing the temperature, a fact that does not seem to have been recognized 
in previous formulations.

As stated in the description of the model, it is necessary to calculate a dynamical property 
(such as the thermal conductivity $\kappa$), 
to get the value of the dynamical exponent $z$.
For finite systems, we can compare the results for a homogeneous lattice 
to those of a non-homogeneously renormalized one, if we assume the effects of 
finite size to be the same.
To determine $\kappa$, several simulations for a 2D homogeneous triangular lattice 
($N_x \times N_y =225 \times 33$) 
and for its homogeneously coarse grained equivalent are run at various temperatures. 
A non-symmetric potential which allows anharmonic phonon interactions, is used.
Since there are no interfaces in the homogeneously coarse grained lattice, 
the only effect is due to the time scaling.
The thermal conductivity $\kappa$ is obtained from the standard Green-Kubo relation 
\cite{ref.kubo-green}. This relation involves time 
integration, so the dynamical exponent $z$ can be implicitly obtained 
by fitting $\kappa$ which is defined per ``unit time''. 
We find that for a value of $z \approx 1.45\pm0.1$, perfect agreement exists for
the $\kappa$ in the atomistic and coarse system.
\begin{figure}
\psfig{file=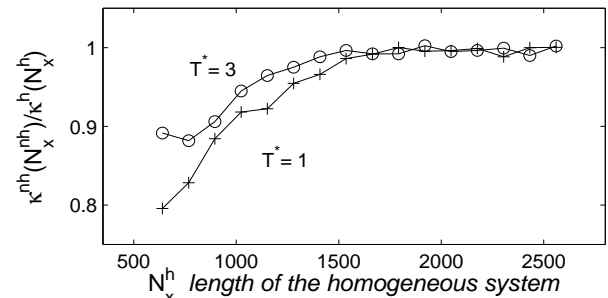,width=80mm,angle=90}
\caption{Interface effect: thermal conductivity ratio $\kappa^{nh}(N^{nh}_x)/\kappa^{h}(N^{h}_x)$ 
        versus the length of a 2D lattice with one interface. $(+)$
	 for $T^\star=1$ and $(\circ)$ for $T^\star=3$. }
       \label{label.fig.3}
\end{figure}

In order to study the thermal transport in non-homogeneous systems
the effect of interfaces between differently coarse-grained regions need to be understood. 
We run simulations for a 2D homogeneous lattice, with $N^{h}=N_x^{h} \times N_y^{h}$ atoms, 
and for its non homogeneous coarse grained version, with $N^{nh}=N_x^{nh} \times N_y^{nh}$
nodes. The non homogeneous lattice has one coarse-graining interface at the center parallel to $y$ direction.
To one side of this interface everything is fully atomistic, 
while to the other side one level of coarse-graining is applied.
Thus $N^{nh}_x=2 N^{h}_x/3$, and each region is $N^{h}_x/3$ wide.
By systematically varying the size of the systems and plotting 
$\kappa^{nh}(N^{nh}_x)/\kappa^{h}(N^{h}_x)$, 
we determine the effect of the interface/size in the thermal conductivity.
Figure \ref{label.fig.3} shows this effect at $T^\star=1$ and $3$:
at high $N^h_x$, the length of the regions becomes comparable to the phonon mean free path.
Hence the fact that there are less nodes in the coarse-grained system
becomes less apparent and $\kappa^{nh}(N^{nh}_x) \rightarrow \kappa^{h}(N^{h}_x)$.
The picture shows that, for our potential, dynamical exponent $z=1.45$, and temperatures ($T^\star=1$ and $3$),
regions wider than $500$ nodes ($N^h_x/3$) give acceptable results. 

\begin{figure}[b]
\psfig{file=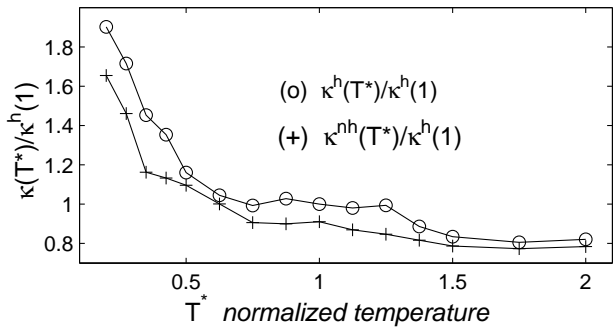,height=45mm,angle=90}
 \caption{Thermal conductivities $\kappa^{h}(T^*)/\kappa^{h}(T^*\!=\!1)\,\,(\circ)$ 
          and $\kappa^{nh}(T^*)/\kappa^{h}(T^*\!=\!1)$ $(+)$ 
          of 2D lattices for various normalized temperatures.}
   \label{label.fig.4}
\end{figure}

For high enough separation between interfaces, the coarse-grained dynamics
reproduces the $T$-dependence of the thermal conductivity well.
Figure \ref{label.fig.4} shows $\kappa^{nh, h}(T^*)/\kappa^{h}(T^* \!=\!1)$
for a 2D lattice with $N^h = N^h_x \times N^h_y = 11264 \times 33 = 371712$ atoms, 
and its non-homogeneous coarse grained version
with $N^{nh} = 63072$ nodes, and 6 interfaces parallel to $y$ axis  
(each region is 512 nodes wide at different level of coarse graining).
The non-homogeneous model underestimates the thermal conductivity
at low temperatures ($T^*<1/3$). 
At high temperatures ($T^*>1$) the two results are within 
$10\sim15\%$.
Normal materials have $T_0$ of the order of room temperature, so the error 
is acceptable considering the strong approximations in the method. 

To conclude, we have proposed and analyzed a molecular dynamics method 
with time and spatial coarse-graining. 
We show that mechanical and thermodynamical properties are in excellent 
agreement with the non coarse grained system. 
If this method can be sufficiently generalized in 3D,
it may form the basis of a RG Molecular Dynamics to investigate effects of
temperature and defects in real nanostructures. 

This research was supported by the Air Force MURI award No. F49620-99-1-0272. 
It has benefited from discussion with Efthimios Kaxiras, Emily Carter, 
Bill Curtin, Ron Phillips, James Sethna, and Dane Morgan.


\begin{thebibliography}{99}
\bibitem{ref.pot.1}  
the renormalized potential does not depend on the temperature if the original potential is harmonic, 
because of the gaussian integration.
\end{thebibliography}
\end{document}